\documentclass[11pt]{article}

\usepackage{amssymb}
\usepackage{amsmath}
\usepackage{epsfig}
\usepackage{fullpage}
\usepackage{graphicx}
\usepackage{array}

\newcommand{\REDUCER}{{\tt REDUCER}}
\newcommand{\ALGOU}{{\tt ALGO2}}
\newcommand{\ALGOD}{{\tt ALGO1}}
\newcommand{\COLORING}{{\tt COLORING}}
\newcommand{\PRESTO}{{\tt PRESTO}}
\newcommand{\BHS}{{\tt BHS}}
\newcommand{\WBHS}{{\tt WBHS}}
\newcommand{\RBS}{{\tt rBHS}}
\newcommand{\ignore}[1]{}

\newcommand{\qed}{\hfill\rule{.5em}{1.5ex}}
\newenvironment{proof}{\begin{trivlist}
    \item[] {\bf Proof:}}{\hspace{1.5em}\qed\end{trivlist}}

\newtheorem{property}{Property}
\newtheorem{theorem}{Theorem}
\newtheorem{corollary}{Corollary}
\newtheorem{lemma}{Lemma}
\newtheorem{fact}{Fact}

\newtheorem{definition}{Definition}

\title{Searching for a dangerous host: randomized vs. deterministic}
\author{Igor Nitto$^*$ and Rossano Venturini\footnote{Department of Computer Science, University of Pisa, Italy.
{\tt \{nitto,rventurini\}@di.unipi.it}} }
\date{}

\begin{document}

\maketitle

\begin{abstract}
A {\em Black Hole} is an harmful host in a network that destroys
incoming agents without leaving any trace of such event. The problem
of locating the black hole in a network through a team of agent
coordinated by a common protocol is usually referred in literature
as the Black Hole Search problem (or \BHS\ for brevity) and it is a
consolidated research topic in the area of distributed algorithms
\cite{inter}. The aim of this paper is to extend the results for
\BHS\ by considering more general (and hence harder) classes of
dangerous host. In particular we introduce {\em rB-hole} as a
probabilistic generalization of the Black Hole, in which the
destruction of an incoming agent is a purely random event happening
with some fixed probability (like flipping a biased coin). The main
result we present is that if we tolerate an arbitrarily small error
probability in the result then the {\em rB-hole Search problem}, or
\RBS, is not harder than the usual \BHS. We establish this result
in two different communication model, specifically both in presence
or absence of whiteboards non-located at the homebase. The core of
our methods is a general reduction tool for transforming algorithms
for the black hole into algorithms for the rB-hole.

\vspace{0.1in} \noindent \textbf{Keywords:} interconnection
networks; malicious hosts; mobile agents; traversal pair;
distributed search
\end{abstract}

\section{Introduction}
The {\em Black Hole Search} problem, or \BHS, \cite{inter,
DobrevKSS06, DobrevFPS06, DobrevFPS01,KMRS07,tree,last} has recently
gained a lot of interest among the research community in mobile and
distributed computation. A Black Hole represents a "malicious" host
in a network, which destroys every agent that tries to pass through
it. No trace of such destruction event will be observable by any
other agent. The \BHS\ problem requires to find a strategy to
coordinate a set of autonomous and mobile agents in order to
discover and correctly report the location of the Black Hole inside
a network. A correct solution is required to terminate after a
finite amount of moves with at least one of the agents surviving and
reporting the correct output.

Several authors have investigated the \BHS\ problem under different
hypothesis about network's topology (like ring \cite{DobrevFPS01},
mesh, hypercube, etc. \cite{inter, last,tree}), kind of
communication devices (i.e., tokens instead of whiteboard
\cite{DobrevKSS06}), network's topological knowledge \cite{last},
presence of multiple black holes \cite{CKR06}, etc.
These different algorithms (or protocols) are usually compared on
the basis of two main complexity measures: the number of moves
performed and the number of agents required, where both this
parameters are taken in the worst case.

In this paper we address the malicious host question in a more
general form, namely we introduce the concept of rB-Hole, which is a
randomized generalization of the Black Hole, and then study new
strategies for its localization in a network. We will see that the
rB-Hole Search problem ( \RBS\ for brevity) problem can be resolved
only if we tolerate an error probability in the output. Under this
hypothesis the \RBS\ problem is solvable and we will provide a
general technique to derive an algorithm for \RBS\ from an algorithm
for \BHS. As a main consequence, by applying our technique to some
of the standard result about the \BHS\ case \cite{DobrevFPS01,
inter}, we provide generalizations of these methods to work for
the \RBS\ problem without increasing asymptotically the number of
moves performed or the number of agents required.

\section{Background and notation}

This section is dedicated to introducing the model of computation
and useful background about the problem. The term {\em agent}
denotes a computational entity allowed to perform an arbitrary
computation. The agents are equipped with a local bounded memory
which maintains the status of their computations or other useful
information. They are able to move themselves in the network by
following the links connecting adjacent nodes. Moreover, the agents
can communicate by reading from and writing on shared memory units
located on the nodes, called {\em whiteboards}. Access to a
whiteboard is done in mutual exclusion. We assume that the amount of
storage available on a whiteboard is $O(\log n)$ bits.

It is important to notice that the agents are {\em asynchronous},
this means there is no assumption on the time taken by an agent to
perform a generic action, like a move on a link or a computation
step. This implies the impossibility to predict when one of this
action will eventually end.

In the following the network will be represented by a connected
undirected graph $G$, whose nodes can be anonymous (i.e., without
unique names).

As defined in \cite{DobrevFPS01}, a {\em black hole} is a stationary
process located at a node, which destroys any agent arriving at that
node. No observable trace of such destruction event will be evident
to other agents. The {\em Black Hole Search} (\BHS) problem
\cite{DobrevFPS01} consists of devising a strategy for coordinating
the agents in order to discover the position of the black hole in a
network.

At the beginning of the strategy all the agents are assumed to be
co-located in a unique safe node called {\em homebase}. After a
finite number of moves, at least one of the agent must survive and
be able to indicate the position of the black hole in the network.

In this paper we propose a generalization of the \BHS\ problem by
introducing the notion of {\em rB-hole}. A rB-hole is an aleatory
process located on an host which can destroy visiting agents with
some fixed probability $p$. More precisely the interaction between a
visiting agent and the rB-hole can be schematized as follows:

\begin{enumerate}

\item The agent move on a link from a safe node to the rB-hole.
The rB-hole flip a biased coin which give HEAD with probability $p$,
where $p$ is a parameter of the rB-hole. If an HEAD comes out the
agent is killed otherwise he advances to the next phase.

\item The agent enters the rB-hole and gains access to its internal
whiteboard. The agent is now safe and hence he is able to consistently
modify the whiteboard.

\item The agent moves on a link from the rB-hole to a safe node.
This phase is symmetrical to the first; a biased coin is flipped and
the agent is killed with probability $p$, otherwise he safely leaves
the rB-hole.

\end{enumerate}

Observe that a black hole is simply a rB-hole with $p$ equal to $1$.
The \RBS\ problem is defined as the analogous of the \BHS\ problem for the rB-hole.

Since the \RBS\ is indeed a generalization of the \BHS\
problem, it automatically inherits all of its known lower bounds.
The following lemmas are therefore immediate corollaries of Lemmas
$1$ and $2$ and Theorem $1$ in \cite{DobrevFPS01}:

\begin{lemma}\label{hardness1}
It is undecidable if a network contains a rB-hole or not using
asynchronous agents.
\end{lemma}

\begin{lemma}
At least two agents are needed to locate the rB-hole.
\end{lemma}

\begin{lemma}
Any algorithm for solving the \RBS\ with asynchronous agents
requires at least $(n-1)\log(n-1) + O(n)$ moves in the worst case.
\end{lemma}

Lemma \ref{hardness1} is perhaps a surprising result, in fact it
implies the impossibility to determine the existence or
non-existence of an rB- (or even black) hole inside a network.
Nevertheless we will give protocols reporting the position of the
rB-hole with arbitrarily small error probability when its existence
is given as hypothesis.

Biconnectedness (or 1-connectedness) of the underlying graph is an
essential hypothesis:

\begin{lemma} \cite{DobrevFPS01}
There are no protocols for the \RBS\ problem over non biconnected
graphs.
\end{lemma}

It's worth to remark that the known protocols for the \BHS\ problem
are not trivially adaptable to the \RBS\ problem since they rely on
the applicability of the {\em cautious walk}\cite{inter,
DobrevKSS06, DobrevFPS06, DobrevFPS01} technique. Cautious walk
consists of leaving information on each traversed nodes in such a
way that an agent can eventually recognize wether a visited link
lead to a safe node or not. The point is that cautious walk strongly
relies on the fact that an agent never survives when he visits the
black hole. Unfortunately this clearly become false in the rB-hole
case as stated in the following:

\begin{fact}
 An agent cannot claim that a traversed node is not the rB-hole.
\end{fact}

In fact we can never ensure that a node is safe after having visited
it. This is actually the main difficulty arising when we pass from
\BHS\ to \RBS. Furthermore, by combining with the obvious fact that
in the asynchronous model it is impossible to distinguish a dead
agent by an agent stuck into a slow link, we get two main
consequences:

\begin{itemize}

\item Classic techniques used in literature for the \BHS\
problem\cite{DobrevFPS01, DobrevKSS06, inter}, like {\em cautious
walks}, are not trivially extendable to the \RBS\ problem.

\item It is always impossible to determine the rB-hole position within a
finite number of moves using asynchronous agents.
\end{itemize}

This last statement seems is rather strong, it basically deny the
existence of a protocol that solves {\em exactly} the \RBS\ problem.
Nevertheless, non-exact solution are still possible; in fact we will
succeed in devising algorithms whose output is correct with
arbitrarily high probability.

The strategies we are going to present relies on two basic
hypothesis:

\begin{enumerate}

\item Every agent has full knowledge about the topology of the graph.

\item A lower bound $p$ on the probability of being
killed by the rB-hole is known by any agent.

\end{enumerate}

In addition our protocols depend on a user specified parameter
$\delta$, which represents an upper bound on the error probability
of the result, i.e., the returned output is correct with probability
at least $1-\delta$.

\section{Traversal pairs}

A {\em traversal pairs} \cite{inter} is a very useful notion for
dealing with \BHS-like problem. We denote by $<_G$ an arbitrary
fixed total ordering $v_1 <_G v_2 <_G \ldots <_G v_n$ of the nodes
of $G$.

\begin{definition}
\textbf{(Traversal Pair)} Let $G = (V,\, E)$ be an $n$-node
biconnected graph with a total ordering $<_G$ of its nodes. Let $\pi
_l$ and $\pi_r$ be two paths on $G$ starting from $u=v_1$ and
$v=v_n$ respectively, and exploring the nodes of $G$ in the order
$v_1,v_2, \ldots, v_n$ and $v_n,v_{n-1}, \ldots, v_1$ respectively.
The pair $\pi = (\pi_l, \pi_r)$ is called $u-v$ traversal pair (TP
for brevity) of $G$ with respect to $<_G$.
\end{definition}

This definition says that if $\pi=(\pi_l,\pi_r)$ is a TP then
starting from $u$ (resp. $v$) and following the path $\pi_l$ (resp.
$\pi_r$) we are able to reach any node $v_j$ of $G$ by crossing only
nodes that are smaller (resp. greater) than $v_j$ in the total
ordering $<_G$.

An $u-v$ traversal pair $\pi=(\pi_l,\, \pi_r)$ of a graph $G$ with
respect to a total ordering $v_{1},\ldots,v_{n}$ of its nodes has
two main parameters: the {\em size} and the {\em radius}.

The size of $\pi$, indicated by $s(\pi)$, is equal to
$max\{|\pi_l|,\, |\pi_r|\}$, where $|\gamma|$ indicates the number
of edges in a generic path $\gamma$. The radius of $\pi$ is defined
by $r(\pi) = max_{w \in V} \{max\{r_u(w),\,r_v(w)\}\}$, where
$r_u(w)$ and $r_v(w)$ are the lengths of the shortest paths that
start from the homebase and reach $w$ by crossing only nodes that
respectively precede or follow $w$ in the total order of $V$.

A graph $G$ is said \textit{traversable} if for any pair of nodes
$u,v \in V$ there exists an ordering $<_G$ of its nodes and an $u-v$
traversal pair with respect to $<_G$. The following is a fundamental
lemma:

\begin{lemma}\label{th:trav} \cite{inter}
A graph $G$ is traversable if and only if it is biconnected.
\end{lemma}

Since the \RBS\ problem is decidable only on biconnected graphs, the
preceding lemma tells us that we can always assume that the input
graph of our algorithm is traversable.

In the following we will assume that the agents share a unique fixed
$(h,v)$-traversal pair $\pi=(\pi_l,\pi_r)$ relative to a total order
$v_1, \ldots, v_n$, where $h$ is the homebase and $v$ is one of its
neighbors. A traversal pair can be constructed from a description of
the graph \cite{inter}, which is provided in our case by the
full-topological knowledge assumption. Therefore, the computation of
$\pi$ can be carried out by every agent before starting the
execution of the protocol and without doing any move.

We use $\pi_l[v_i,v_j]$ to indicate the subpath of $\pi_l$
connecting the first occurrences of $v_i$ and $v_j$ in $\pi_l$.
Analogously do for $\pi_r[v_i,v_j]$. The notation $[v_i,v_j]$
indicates the subset $\{v_i,\ldots,v_j\}$ of the nodes, we will use
the term {\em interval} to refer to one such subset.

\section{From black to rB-hole}

We are ready to derive our first protocol for the \RBS, which
requires the presence of whiteboard on each node. We will present
our result using a reduction paradigm. In fact we will provide a
general methodology to extend a protocol for \BHS\ into one for
\RBS. The main tool exploited here is a particular "coloring"
protocol. It requires two agents, which explore the nodes along the
two directions of the common traversal pair assigning colors to
them.

This protocol warrantees that if one of the agents completes its
execution it can report the rB-hole location with arbitrarily high
probability, otherwise, if both the agents die, then the rB-hole,
together with at most a constant number of nodes, is marked with a
different color. In both of these scenarios we have enough
information to solve the problem. In fact we will see in subsection
\ref{aftercoloring} how to combine the coloring protocol and a generic
protocol for \BHS\ to obtain a corresponding protocol for \RBS; in
the most notable cases (such as algorithm \PRESTO\ \cite{inter}) the
resulting algorithms will have the same complexity in term of number
of moves and agents.

\subsection{Protocol \COLORING}

Protocol \COLORING\ requires two agents, say $a^l$ and $a^r$, which
perform a visit of all nodes starting respectively from nodes $v_1$
and $v_n$ and following respectively paths $\pi_l$ and $\pi_r$.
Recall that $v_1, \ldots v_n$ is the ordering of the nodes given by
the chosen traversal pair. The agents leave information on the
whiteboard of every touched node in order to encode the number of
times they have traversed it. The information associated to a
generic node $v$ is actually a value in $\{0,1,2,3\}$, for
simplicity we think it as the {\em color} of $v$ and denote it by
$c(v)$.

Let us describe the actions taken by $a^l$ and $a^r$ whenever they
enter a node $v_i$ along their path. The following list shows the
behavior of the generic agent $a^{id}$ according to the color of
$v_i$. We set the {\em predecessor} of node $v_i$ respectively equal
to node $v_{i-1}$ if $id = l$ or $v_{i+1}$ if $id = r$. The constant
$\Delta$ will be equal to $ \lceil\log_{1-p} \delta \rceil$, where
$0 < \delta \leq 1$ is the user defined probability error.

\begin{enumerate}

\item If $c(v_i) = 0$ then $v_i$ is unexplored and $a^{id}$ sets $c(v_i)=1$ and moves back 
    on $\pi_{id}$ to set the color of the predecessor of $v_i$ to $3$.
    After that, $a^{id}$ return to $v_i$ and moves back and forth $\Delta$ times from the last visited
    node. Such behavior will be referred from now on as a {\em $\Delta$-visit} of a node.
    Once the $\Delta$-visit of $v_i$ eventually ends (i.e. $v_i$ is not the rB-hole), $a^{id}$ 
    set $c(v_i) = 2$ and continues exploring path $\pi_{id}$.

\item If $c(v_i) = 1$ or $c(v_i) = 2$ and $v_i$ has never been visited before by $a^{id}$ then $a^{id}$
moves back to the homebase indicating $v_i$ as the rB-hole.

\end{enumerate}

The information encoded in the color of a node in this protocol has
the following meaning. Initially we assume that the color of every
unexplored node is $0$ which correspond to an empty whiteboard.
Successively a node can be colored first $1$ and then $2$ when it is
respectively visited for the first or for the $\Delta$-th times.
Moreover, for any $i \in [n]$, $c(v_i)$ is set equal to $3$ by agent
$a^{l}$ (resp. $a^{r}$) iff $v_{i+1}$ has been visited at least one
time (resp. $v_{i-1}$).

The execution of this protocol may have two possible outcome. In
fact either the protocol terminates with one of the agents at the
homebase reporting an output, or the protocol might fail, in which
case both of the agents are killed by the rB-hole. Notice that both
of the agents may terminate their execution with two distinct
reported node, in this case we break the ties choosing the output
reported by the first agent reaching the homebase.

Now we outline the main property of \COLORING. First of all this
protocol can be interpreted as a semi-solution for the \RBS. This
simply means that, as shown in the next theorem, when it terminates
its output is correct with high probability:

\begin{lemma}\label{correct}
If \COLORING\ terminates then the outputted node is the rB-hole with
probability at least $1-\delta$
\end{lemma}

The only problematic case is when the protocol fails to terminate.
However, in this case we can still exploit the information coming
from the coloring in order to reduce to a constant the number of
nodes that possibly contain the rB-hole.

\begin{lemma}
Let $g$ be the index of the rB-hole and suppose that the protocol
does not terminate. After the destruction of both the agents we have
that:
\begin{enumerate}
 \item $c(v_g) < 3$
 \item $c(v_j) = 3$ for any $j \neq g-1,g,g+1$, and $c(v_{g-1})
 \geq 2$, $c(v_{g+1}) \geq 2$.
\end{enumerate}

\end{lemma}

It remains to observe that the number of moves performed by coloring
is at most $O(\Delta n + s(\pi))$. In fact it takes $O(s(\pi))$
moves to traverse $\pi$ plus another $O(\Delta n)$ moves to
$\Delta$-visit each node.

\subsection{Turning a \BHS\ protocol into a \RBS\ protocol}\label{aftercoloring}

In this section we show how to combine a standard \BHS\ protocol
with the \COLORING\ protocol of the previous subsection in a unique
solution for \RBS. Let $A$ be a correct asynchronous protocol for
\BHS\ (i.e. \PRESTO \cite{inter}). 
We assume that the original protocol always terminates with
at least one agent surviving and reporting the right output in the
homebase.

We are going to define two small variant $A_0$, $A_1$ of $A$. In
particular protocol $A_j$ is equal to protocol $A$ with the
difference that an agent suspends its task and saves its internal
status whenever he enters on a node $v_i$ such that one of the following
two {\em virtual black-hole} conditions is verified:

\begin{enumerate}
 \item $i \equiv j$ (mod 2), and $c(v_i) < 3$.
 \item $i \equiv j+1$ (mod 2), and $c(v_i) < 2$.
\end{enumerate}

Whenever in a later time this conditions become false the agent
involved will restores the last status before stopping and continues
with the execution of the protocol.

The following lemma states a key property:

\begin{lemma}\label{termination}
Let $v_g$ be the node containing the rB-hole and consider a parallel
execution of \COLORING, $A_0$ and $A_1$. If \COLORING\ fails to
terminates and $g \equiv k$ (mod 2) then $A_k$ terminates reporting
$v_g$ as output.
\end{lemma}

Therefore at least one among \COLORING, $A_0$ and $A_1$ will
terminate. We already know from lemma \ref{correct} that \COLORING\ reports
a probably correct output whenever it terminates. The
following lemma establishes the same result for $A_0$ and $A_1$:

\begin{lemma}\label{reliability}
Let $v_g$ be the node containing the rB-hole and consider a parallel
execution of \COLORING\ and $A_j$, for any $j\in\{0,1\}$. The
probability that $A_j$ reports $v_g$ as rB-hole given that its
execution terminates is at least $1-\delta$.
\end{lemma}

By combining lemmas \ref{termination} and \ref{reliability} we are
now ready to exhibit the main result of this section:

\begin{theorem}\label{mainth}
Let $A$ be an asynchronous protocol for the \BHS\ problem on a
network $G$ of $n$ nodes, requiring at most $t$ agents and $m$ moves
and let $\pi$ be a TP of $G$. For any $0 < \delta \leq 1$ there
exists a protocol $A_r$ for \RBS\ on $G$ requiring at most $2t + 2$
agents and $O(m+s(\pi)+\Delta n)$ moves, where $\Delta = \log_{1-p}
\delta \thickapprox O(\frac{\log 1/\delta}{p})$. $A_r$ always
terminates and reports the correct output with probability at least
$1-\delta$.
\end{theorem}

To further understand the power of theorem \ref{mainth} let us
mention two corollaries obtained instantiating $A$ with optimal
\BHS\ protocols. In particular using the results in \cite{inter} we
obtain:

\begin{corollary}\label{general}
For any $0 < \delta \leq 1$ there exists a protocol for \RBS\ on
general network requiring $O(1)$ agents and $O(s(\pi) + \Delta n +
r(\pi)\log r(\pi))$ moves where $\Delta \thickapprox \frac{1}{p}
\log(1/\delta)$. The protocol always terminates and the reported
output is correct with probability at least $1-\delta$.
\end{corollary}

Finally, as a last example of our technique, we derive a version of
theorem \ref{mainth} for ring topology:

\begin{corollary}\label{ring}
The \RBS\ problem can be solved on an $n$-nodes ring with
arbitrarily high constant probability using $O(1)$ agents and
$O(n\log n)$ moves
\end{corollary}

This results actually say that solving the \RBS\ with an arbitrarily
small error probability is not harder than solving \BHS. In fact
theorem \ref{mainth} provides a \RBS\ protocol having the same
asymptotical complexity of the best known \BHS\ protocol for general
network \cite{inter}, if the error probability is considered as a
constant. In the case of ring topology it has been proved
\cite{DobrevFPS01} that at least $O(n\log n)$ moves are necessary to
solve \BHS\ with $O(1)$ agents, hence corollary \ref{ring} is also
optimal.

\section{A strategy for \RBS\ in absence of whiteboards}

In this section we continue the analysis of the \RBS\ problem by
considering a more restrictive communication model. Namely we will
show how to solve \RBS\ without making use of whiteboards non
located in the homebase.

We use $\pi_l[v_i,v_j]$ to indicate the subpath of $\pi_l$
connecting the first occurrences of $v_i$ and $v_j$ in $\pi_l$.
Analogously do for $\pi_r[v_i,v_j]$. The notation $[v_i,v_j]$
indicates the subset $\{v_i,\ldots,v_j\}$ of the nodes, we will use
the term {\em interval} to refer to one such subset. We need to
define a weight function $w$ for assigning to any interval the
corresponding amount of moves needed to visit it in both directions
of $\pi$. We do this by taking $w([v_i,v_j])= |\pi_{l}[v_i,v_j]| +
|\pi_{r}[v_i,v_j]|$. An interval $[v_q,v_z]$ will be called {\em
viable} when $w([v_q,v_z]) \leq 6r(\pi)$ and $|[v_q,v_z]| \leq
r(\pi)$.

We require for the traversal pair $\pi$ to satisfy the following
useful property:

\begin{property} \label{peso}
For any node $v_i$ we have $w([v_i,v_{i+1}]) \leq 4r(\pi)$.
\end{property}

Indeed this is not a restriction, in fact, for the definition of
radius, it is always possible to replace $\pi_{l}[v_i,v_{i+1}]$ or $
\pi_{r}[v_{i},v_{i+1}]$ with a path of length at most $2r(\pi)$,
when necessary, and preserving the consistency of the traversal
pair. In this way we can always force a traversal pair to satisfy
Property \ref{peso}.

\subsection{First step: reducing the problem to viable intervals} \label{sreducer}

The first step of our solution consists of finding a viable interval
of nodes containing the rB-hole with arbitrarily high probability.

To this aims we define a protocol called \REDUCER\ which requires
two agents, named $a^l$ and $a^r$. The agents compute a partition
${\cal L} = \{U^1, \ldots, U^f \}$ of the nodes $V=[v_1, v_n]$ in
viable intervals that respects the following constraints:

\begin{enumerate}
 \item $\bigcup _{i=1} ^f U^i= V$
 \item For any $i$ and $j$, $U^i \cap U^j = \emptyset$
 \item For every $1 \leq i < f$ we have $2r(\pi) \leq w(U^i)$.
\end{enumerate}

The existence of ${\cal L}$ is ensured by Property \ref{peso} and it
can be computed by the agents without performing any move
 because of the full-topological knowledge assumption.

After that, $a^l$ and $a^r$, start to explore the intervals in
${\cal L}$ according to the following strategy, which guarantees
that any interval is explored by at most one agent. The agents $a^l$
and $a^r$ respectively explore the intervals in increasing or
decreasing order of index starting from $1$ or $f$. Every time the
visit of an interval terminates, the agent comes back to the
homebase, writes on the whiteboard the index of the last visited
interval and decides whether or not to start the visit of the next
interval. The protocol ends when at least one of the agents realizes
that only one interval is left to explore. This interval is claimed
to contain the rB-hole. Put $\Delta= \lceil\log_{1-p} \delta
\rceil$, where $0 < \delta \leq 1$ is an user defined parameter. The
following procedure is performed by $a^l$ and $a^r$ in order to
visit a generic interval $U^i=[v_p, v_q]$, with $id$ instantiated
respectively to $l$ or $r$ according to the identity of the agent:

\begin{enumerate}
 \item Move from the homebase to $v_p$  using a path of
       length at most $r(\pi)$ that crosses only nodes in
       $[v_1, v_p]$ if $id = l$ or $[v_p, v_n]$ otherwise.
 \item Traverse $\pi_{id}[v_{p},v_q]$, every time a nodes
       $u \in U^i$ is visited for the first time move
       back and forth from $u$ to the previously visited node for $\Delta$ time.
       In the following we refer to the latter behavior as a
\textit{$\Delta$-visit} of $U^i$.
 \item Move from $v_q$ to the homebase via a path of
       length at most $r(\pi)$ that crosses only nodes in
       $[v_1, v_q]$ if $id = l$ or $[v_p, v_n]$ otherwise.
 \item Write on the homebase whiteboard that $U^i$ has been visited.
\end{enumerate}

Notice that $\Delta$-visiting is crucial in order to increase the
probability to be destroyed by the rB-hole. It is easy to see that
there does not exist a node which is visited by both agents. Thus,
\REDUCER\ always terminates with at least one agent survived and
reports an interval $U^o$ as output.

\begin{lemma}\label{reducer}
Protocol \REDUCER\ requires two agents and $O(s(\pi) + \Delta n)$
moves, where $\Delta=\lceil\log_{1-p} \delta\rceil \simeq
\frac{1}{p}\log \frac{1}{\delta}$ and $0 < \delta \leq 1$ is an user
defined error probability, and returns a viable subinterval
containing the rB-hole with probability greater than $1 - \delta$.
\end{lemma}

\begin{proof}
It is easy to see that two agents are enough to complete the
protocol. By the properties of ${\cal L} = U^1, \ldots, U^f$ and by
our choice of weight function $w$ it follows that $s(\pi)/3r(\pi)
\leq f \leq s(\pi)/r(\pi)$. Therefore, we can reach the first node
of each interval and come back to the homebase with
$2r(\pi)f=O(s(\pi))$ moves. We can observe that the number of moves
required to traverse every interval is bounded by $\sum _{i=1} ^f
w(U^i)\leq 2s(\pi)$. Instead, since the intervals are pairwise
disjoint, we $\Delta$-visit any node at most one time. Thus, the
overall number of moves is $O(s(\pi)+\Delta n)$.

Exactly one of $a^l$ and $a^r$ will $\Delta$-visit the interval $I
\in {\cal L}$ containing the rB-hole. The reported output $U^o$ can
be different from $I$ only if the agent that visits $I$ survives.
Since he will traverse the rB-hole at least $\Delta$ time, the
probability that the protocol reports a wrong output is upper
bounded by $(1-p)^\Delta \leq \delta$.
\end{proof}

\subsection{Second step: searching inside a viable interval}\label{reduced}

In the second step of our solution we will discover the position of
the rB-hole inside the viable interval $U = [v_q, v_z]$ provided by
the first step. If $U$ truly contains the rB-hole then the returned
output is correct with probability greater than $1-\delta$, where
$\delta$ is an arbitrarily small constant defined by the user.

Here we present two different protocols for this subproblem,
offering a trade-off between number of moves and number of agents.
The first one, called \ALGOU, works in $O(r(\pi)^2)$ moves using
only 2 agents while the second, \ALGOD, requires $O(r(\pi)\log
r(\pi))$ moves and $\rceil \log r(\pi) + 1 \lceil$ agents (recall
that size and weight of the viable interval are in $O(r(\pi))$).

\ALGOD\ takes the viable interval $U$ as input and mimic a binary
search over its nodes. This consists of a sequence of {\em stages},
where the generic $t$-th stage has associated a subinterval $V^t$ of
$U$. Initially we put $V^{1}=U$. This is successively splitted stage
by stage until it contains only one node, which is reported as the
rB-hole.

Now we describe the computations performed at the generic $t$-th
stage over $V^t = [v_l,v_r]$. We choose $k = \lfloor(r+l)/2\rfloor$
as a pivot to partition $V^{t}$ into two disjoint subintervals,
$V^t_{l} = [v_l,v_k]$ and $V^t_{r} = [v_{k+1}, v_r]$. Then we select
two available agents at the homebase, say $b^{l}$ and $b^{r}$, which
execute in parallel the following list of actions, with $id$
respectively equal to $l,r$:

\begin{enumerate}

\item Put $q = k$ if $id=l$ or $k+1$ otherwise

\item Move from the homebase to $v_{id}\in V^t_{id}$ via a path of length at
most $r(\pi)$ that crosses only nodes in $[v_1, v_{id}]$ if $id = l$
or $[v_{id}, v_n]$ otherwise.

\item Traverse $\pi_{id}[v_{id},v_q]$ and $\Delta$-visit $V^t_{id}$ with
      $\Delta = \lceil\log_{1-p}(\delta/(\log r(\pi) + 1))\rceil$.

\item Move from $v_q$ to the homebase via a path of length at most
$r(\pi)$ that crosses only nodes in $[v_1, v_q]$ if $id = l$ or
$[v_q, v_n]$ otherwise.

\end{enumerate}

Stage $t$ terminates when one of the two agents first completes its
task. $V^{t+1}$ is set equal to $V^t_l$, if $b^r$ terminates its
task before $b^l$, or to $V^t_r$ otherwise. This operation requires
the use of the solely homebase whiteboard: when one among $b^l$ and
$b^r$ comes back to the homebase he reads the whiteboard to discover
if he has been the earlier, then he eventually updates the status of
the protocol by setting the parameters of the next stage.

Notice that every stage terminates within a finite amount of time.
In fact the sets of nodes visited by $b^l$ and $b^r$ are disjoint,
thus only one of them can contain the rB-hole and at least one of
the agents survives and comes back to the homebase.

\begin{lemma}\label{sol2}
\ALGOD\ requires $O(r(\pi)\log r(\pi) + \Delta r(\pi))$ moves and
$\log r(\pi) + 1$ agents, where
$\Delta=\lceil\log_{1-p}(\delta/(\log r(\pi) + 1))\rceil \simeq
\frac{1}{p}\log\left( \frac{\log r(\pi)+1}{\delta} \right)$, and
returns the correct rB-hole position with probability at least
$1-\delta$, where $0<\delta \leq 1$ is an user defined parameter.
\end{lemma}

\begin{proof}
By definition of viable interval $|U|\leq r(\pi)$ and by the fact
that in each stage we (almost) halve the size of the interval, it
follows that the number of stages is at most $\log r(\pi)$. Since in
each of them we lose at most one agent, the number of agents
required by \ALGOD\ is at most $\log r(\pi)+1$.

In the generic stage $t$, the moves performed by the agents are
classified into:
\begin{enumerate}
 \item Moves for reaching the first vertex in $V^t$.
 \item Moves for traversing and $\Delta$-visiting each node in $V^t$.
 \item Moves for reaching the homebase.
\end{enumerate}

By definition of radius, points 1 and 3 require $O(r(\pi))$ moves
per stage. Thus, over all stages these require $O(r(\pi)\log
r(\pi))$ moves. As far as point 2 is concerned, we notice that the
number of nodes $\Delta$-visited in $t$-th stage is equal to
$|V^t|$. Since the size of this intervals decreases geometrically,
the $\Delta$-visits do not require more than $O(\Delta r(\pi))$
moves. We also observe that for every stage $t$, $w(V^t)\leq w(U)$
is smaller than $6r(\pi)$. Since the weight bounds the number of
moves to visit an interval, the number of moves required to traverse
all the intervals is $O(r(\pi)\log r(\pi))$. Summarizing, the whole
algorithm requires $O(\Delta r(\pi) + r(\pi)\log r(\pi))$ moves.

Our protocol fails to indicate the rB-hole if in one of the stage an
agent survives even though it has the rB-hole in its interval. In
any stage this happens with probability less than $(1-p)^\Delta \leq
\delta/(\log r(\pi) +1)$. Using the Union Bound over all the stages
of the protocol we conclude that the final output is not correct
with probability less than $\delta$.
\end{proof}

\ALGOU\ is a slight variant of \REDUCER. We will not give a
detailed description, since the only difference with \REDUCER is
that we have single nodes rather than viable interval to be
validated in each round by the agents. Using these ideas we can
prove the following:

\begin{lemma}\label{sol1}
 \ALGOU\ requires $O(r(\pi)^2 + \Delta r(\pi))$
 moves and $O(1)$ agents, where $\Delta = \log_{(1-p)} \delta$. The
 returned output is the rB-hole with probability at least $1-\delta$, where $0 < \delta \leq 1$ is an user defined
 parameter.
\end{lemma}

\subsection{Summarizing}\label{final}

If we combine \REDUCER\ (subsection \ref{sreducer}) with
respectively \ALGOD\ and \ALGOU\ (subsection \ref{reduced}) we
derive the following two results:

\begin{theorem}\label{comp1}
There exists a protocol for \RBS\ problem that requires $O(1)$
agents and $O(s(\pi) + r(\pi)^2 + \Delta n)$ moves, where
$\Delta=\lceil\log_{1-p} \delta \rceil \simeq \frac{1}{p}\log
\frac{1}{\delta}$ and $0<\delta \leq 1$ is an user defined
parameter. At least one agent survives and fails to indicate the
rB-hole with probability less than $\delta$.
\end{theorem}

\begin{theorem}\label{comp2}
There exists a protocol for \RBS\ problem that requires $\lceil \log
r(\pi)+2 \rceil$ agents and $O(s(\pi) + \Delta n + r(\pi)\log
r(\pi))$ moves, where $\Delta=\lceil\log_{1-p} (\delta/(\log r(\pi)
+ 1))\rceil \simeq \frac{1}{p}\log\left( \frac{\log
r(\pi)+1}{\delta} \right)$ and $0<\delta\leq 1$ is an user defined
parameter. At least one agent survives and fails to indicate the
rB-hole with probability less than $\delta$.
\end{theorem}

Finally, merging together the last two theorems, we establish the
main result of this section:

\begin{theorem}\label{comp3}
For any $0 < \delta \leq 1$ there exists a protocol for \RBS\
problem that requires:
\begin{enumerate}
\item  $O(1)$ agents and $O(s(\pi) + \Delta n + r(\pi)^2)$ moves, where
$\Delta=\lceil\log_{1-p} \delta\rceil \simeq \frac{1}{p}\log
\frac{1}{\delta}$
\item $\lceil \log r(\pi)+2 \rceil$ agents and $O(s(\pi) + \Delta n + r(\pi)\log
r(\pi))$ moves, where $\Delta=\lceil\log_{1-p} (\delta/(\log r(\pi)
+ 1))\rceil \simeq \frac{1}{p}\log\left( \frac{\log
r(\pi)+1}{\delta} \right)$, otherwise
\end{enumerate}
At least one agent survives and fails to indicate the rB-hole with
probability less than $\delta$.
\end{theorem}

Since \RBS\ is a generalization of \BHS, our protocol is also
suitable for the \BHS\ problem. In fact, if the rB-hole is a black
hole (i.e. $p = 1$) our protocol always terminates with the correct
answer using $\Delta = 0$. Therefore, by Theorem \ref{comp3} it
follows:

\begin{corollary}
There exists a protocol for \BHS\ that requires
\begin{itemize}
 \item $O(1)$ agents and $O(s(\pi))$ moves, if $r(\pi)=O(\sqrt{s(\pi)})$
 \item $O(\log r(\pi))$ agents and $O(s(\pi) + r(\pi)\log r(\pi))$ moves, otherwise
\end{itemize}
and uses a unique whiteboard in the homebase.
\end{corollary}

On the other hand, if the use of additional communication devices is
forbidden, we cannot hope to significantly reduce the number of
agents without increasing the number of moves. This is a consequence
of the following theorem whose proof sketch is deferred to the
appendix:

\begin{theorem} \label{lowbound}
 Any protocol which solves the \BHS\ problem for any biconnected graph
 using only one whiteboard and performing at most $O(s(\pi) + r(\pi)\log r(\pi))$ moves must
 use $\Omega(\log r(\pi))$ agents in the worst case.
\end{theorem}

\section{Conclusions}
In this paper we have introduced and studied the rB-hole Search
(\RBS) problem as a probabilistic generalization of the Black Hole
Search problem. We have provided a protocol that solves it with
arbitrary high probability. Even being a generalization of the \BHS\
problem, our protocol for \RBS\ requires asymptotically the same
number of moves and only a slightly larger amount of agents with
respect to \BHS's protocols. We have provided provided solutions
under different communication model.

\bibliographystyle{plain}
\bibliography{grey}

\appendix
\section{Proof sketch for Theorem \ref{lowbound}}

We will actually prove this lower bound under even stronger
hypothesis. Consider a variant of the black hole that {\em marks}
the agents instead killing them and assume that it is possible to
discover whether an agent is marked or not only when he is located
in the homebase. Let us call \WBHS\ the relative searching problem.
It is almost trivial to see that \WBHS\ is simpler than \BHS, i.e:
a lower bound for \WBHS\ must be valid for \BHS\ too.

Having this in mind we prove by contradiction that any protocol $P$
that solves \WBHS\ on an $n$-nodes ring $R$ with only one whiteboard
and using less than $C n\log n$ moves in the worst case must use at
least $c' \log n$ agents for sufficiently large $n$, where $c'$ is a
constant depending only on $C$. Since the radius and the size of the
traversal pair of a ring are both linear this implies the theorem.

Since every agent comes back to the homebase after a finite time, we
can assume w.l.o.g. that in every instant at most one of the agents
is located outside the homebase and, starting from the homebase,
perform a visit of a subset of consecutive nodes before coming back
to the homebase. Assume a numbering of the nodes consistent with the
clockwise order. We will describe a visit as a couple $<E,c>$, where
$E$ is the set of node visited and $c$ is equal to $1$ if the black
hole belong to $E$ or $0$ otherwise.

Put $c' = min\{1/25, 1/log(12C), 2C\}$ and admit by absurd that at
most $m < c'\log n$ agents are marked in the worst case during an
execution of $P$. Denote by $F$ the set of nodes at distance at
least $n/4$ by the homebase. Consider the set of all \WBHS\
instances $I$ having the black hole in $F$. Let $<E_1,c_1> \ldots,
<E_r,c_r>$ be the sequence of all visits of $P$ that touch at least
one node in $F$, ordered by starting time. Observe that $r \leq
4C\log n$ since $|E_j| > n/4$ for any $1\leq j \leq r$ and $P$
perform at most $Cn\log n$ moves. In addition the output of an
execution of $P$ on an instance in $I$ is uniquely determined by the
bit sequence $c_1 \ldots c_r$, and this sequence can contain at most
$m$ ones, since at most $m$ agents can be marked during any
execution of $P$. This implies that the number of distinct output of
$P$ on instances in $I$ is upper bounded by the number of $4C\log
n$-length bit string with at most $m$ ones. But the number of
distinct outputs must be at least $|F|>n/2$ therefore we must have

\[ c'\log n {4C\log n \choose c'\log n} \geq \sum _{j=1}^{m} {4C\log n \choose j} \geq
n/2 \]

\noindent which is impossible for sufficiently large $n$ by our
choice of $c'$.

\end{document}